\documentstyle[multicol,prl,aps,epsf,epsfig]{revtex}

\begin{document}

\title{Material-specific spin filtering in ferromagnet/superconductor ballistic
nanojunctions}
\author{F. Taddei\thanks{e-mail: Fabio.Taddei@dmfci.ing.unict.it},}
\address{School of Physics and Chemistry, Lancaster University,
Lancaster, LA1 4YB, UK\\
Dipartimento di Metodologie Fisiche e Chimiche (DMFCI), Universit\`a di Catania,
Viale A. Doria, 6, I-95125 Catania, Italy\\
Institute for Scientific Interchange (ISI) Foundation, Viale Settimio Severo, 65,
I-10133 Torino, Italy}
\author{S. Sanvito\thanks{e-mail: ssanvito@mrl.ucsb.edu},}
\address{Materials Department, University of California, Santa Barbara,
CA 93106, USA}
\author{C. J. Lambert}
\address{School of Physics and Chemistry, Lancaster University,
Lancaster, LA1 4YB, UK}
\date{\today}
\maketitle

\begin{abstract}
We study spin-dependent electronic transport across ferromagnet/superconductor
ballistic junctions modeled using tight-binding Hamiltonians with s, p and d
orbitals and material-specific parameters. We show that by accurately modeling
the band structure of the bulk materials, one can reproduce the measured
differential conductance of Cu/Pb nanocontacts \cite{Upd1,Upd2}. In contrast
the differential conductance of Co/Pb contacts can only be reproduced if an
enhanced magnetic moment is present at the interface.
\end{abstract}

\begin{multicols}{2}

\vspace{5mm}

During the last few years numerous experimental studies of
electronic transport properties of nanostructures containing both
ferromagnets (F) and superconductors (S) have been reported
\cite{Upd1,Upd2,Soul,Fierz,Gior1,vasco,Pann,Petra,Gior2,jed,Bour}.
Such structures exhibit novel features, not present in
normal-metal/superconductor (N/S) junctions, due to the
suppression of electron-hole correlations in a ferromagnet when a
large exchange field is present. Spin-dependent transport in
structures containing magnetic materials is also underpinning
technological advances in spintronics, where magnetic materials
are used as spin-filters. A key parameter is the degree of
polarization $P$ of the current in a ferromagnet, which is
currently the subject of an intense debate (see, for example
\onlinecite{Fert,Maz,Nadg,Alt}). In the two spin fluid approach
$P$ is defined as:
\begin{equation}
P=\frac{I^\uparrow - I^\downarrow}{I^\uparrow + I^\downarrow} ~,
\label{pol}
\end{equation}
where $I^\uparrow (I^\downarrow)$ is the current carried by
spin-up (spin-down) electrons. Unfortunately $I^\uparrow$ and
$I^\downarrow$ cannot be measured separately in an isolated
ferromagnet and therefore $P$ cannot be determined directly. As
Tedrow and Meservey showed in references
\onlinecite{TM1,TM2,Para,TM3}, $P$ can be estimated by attaching F
to a superconductor through a tunnel junction and taking advantage
of the superconducting gap in the density of states (DOS) of the
superconductor. This
method, however, has a limitation, namely that the insulating
layer has to be uniform, which is a difficult situation to reach
for many ferromagnetic materials. In particular atomic size
pin-holes can short-circuit most of the tunneling current
\cite{Schuller}, giving rise to spurious $I$-$V$ tunneling curves.

To overcome this problem, an alternative method has been proposed
\cite{Upd1,Upd2,Soul} which exploits the suppression of Andreev
reflection at F/S ballistic junctions. In this Letter we argue
that an understanding of spin-polarized transport in such hybrid
nanostructures requires an understanding of surface scattering
which goes beyond the heuristic analysis of references
\cite{Upd1,Upd2,Soul}. We present detailed calculations of the
conductance of Co/Pb and Cu/Pb ballistic interfaces, which show
that although the experiments of references \cite{Upd1,Upd2} tell us
little about bulk magnetization, they do provide the first
evidence of enhanced surface magnetism at the Co/Pb interface.

Providing that S is much longer than the superconducting coherence
length, the sub-gap conductance of a F/S junction is solely
determined by Andreev reflection at the interface. The idea used
in references \onlinecite{Upd1,Upd2,Soul} to estimate the
polarization is based on the fact that, in the absence of
spin-flip processes, as $P$ is increased, Andreev reflection is
suppressed in favor of normal reflection. In the present
calculation, F/S and N/S junctions are described using a
tight-binding Hamiltonian on a $fcc$ lattice with hopping to first
nearest neighbors. In order to accurately reproduce the band
structure of real materials, we take into account 9 orbitals per
site (s, p and d) and calculate the tight-binding parameters by
fitting the band structure obtained from density functional
calculations \cite{Papa.6}. The fit is made using OXON
\cite{OXON}, a tight-binding code which minimizes the deviation
between the LDA results and dispersion curves obtained from the
tight-binding parameterization. As reference points in the band
structure, we take the eigenvalues at four high symmetry points in
the $fcc$ Brillouin zone (namely $\Gamma$, $L$, $X$ and $W$) of
each band. Moreover we further checked the symmetry of the
resulting tight-binding bands along several directions in the
Brillouin zone. It is worth noting that, in order to get a good
fit for both majority and minority electrons of Co, the band
structures of the different spin species are fitted separately as
if they were different materials.

The junction is modeled by coupling a ferromagnetic semi-infinite
lead on the left-hand-side to a superconducting semi-infinite lead
on the right-hand-side, using an interface Hamiltonian $H_{int}$.
The hopping matrix elements are chosen to be the mean square of
the bulk elements, with a sign equal to that of the largest of the
two bulk parameters.\footnote{ It should be noted that other
choices have been considered in the literature, including the
geometric mean using the above sign rule and the geometric mean
with a sign equal to that of the product of the bulk parameters.
We have repeated the calculations of this Letter for both of these
choices and find that neither is capable of reproducing the
experimental results for both Co/Pb and Cu/Pb.}
The conductance of the junction is evaluated within the scattering approach
outlined in \onlinecite{Ste} where transport amplitudes are calculated using
a recursive Green's function technique.

Since we consider only clean interfaces the system is
translationally invariant in the directions parallel to the
interface, so that the total scattering coefficients are given by
the sum over all possible Bloch wave-vectors in the 2-dimensional
Brillouin zone. In our calculations we take the interface
perpendicular to the (110) direction and sum over 900 Bloch
wave-vectors, which corresponds to a junction diameter of the
order of the experimental one (junction area $\sim$10~nm$^2$). In
order to make a comparison with the experimental data
\cite{Upd1,Upd2}, we define the following dimensionless quantity:
\begin{equation}
g(V)= \frac{G_S(V) - G_N(V)}{G_N(0)} \; {,}
\label{g}
\end{equation}
where $G_S(V)$ ($G_N(V)$) is the differential conductance at
voltage $V$, when the S-lead is in the superconducting (normal) state.

\begin{figure}[h]
\narrowtext
\epsfysize=6cm
\epsfxsize=8cm
\centerline{\epsffile{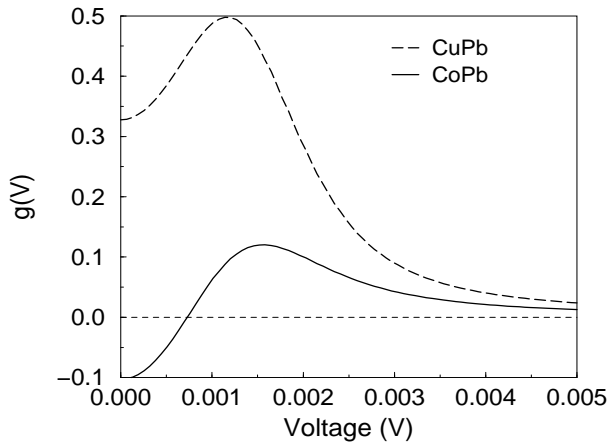}}
\caption{\it \label{g_averages} $g(V)$ curves obtained using the spd-model. The
dashed line is relative to the Cu/Pb junction, while the solid line is relative to
the Co/Pb junction.}
\end{figure}

In Fig. \ref{g_averages} the computed $g(V)$ curves are plotted
for Cu/Pb and Co/Pb junctions at T=4.2 K using the superconducting
gap for bulk Pb ($\Delta=1.26$ meV). Fig. \ref{g_averages} shows
that the measured $g(V)$ curve of references
\onlinecite{Upd1,Upd2} is well reproduced for Cu/Pb, but in the
absence of surface magnetism, the Co/Pb result disagrees with
experiment. At present there exist no ab-initio calculations of
the Co/Pb interface, mainly due to difficulties associated with
modeling heavy elements such as Pb. Furthermore little is known
experimentally about surface magnetism at this important
interface. Nevertheless it is known that related interfaces can
yield surprises. For example in recent experiments involving F/S
multilayers and F/S trilayers both the presence and absence of
magnetically dead monolayers at the surface, when S is in the
normal state, have been reported depending on the material and
geometry (see for example \onlinecite{muhge1,muhge2,garif}). In
addition, an enhanced magnetic moment has been found in
ferromagnetic clusters, isolated or deposited onto a film. For
exampler in experiments on Co, Ni and Fe clusters
\cite{clu1.1,clu1.2,clu1.3}, the magnetic moment was found to
increase up to 36\% higher than the bulk value. LMTO calculations
of Co islands grown on Cu films \cite{clu2} also show an increase
of about 40\% in the local spin polarization.
This enhancement of the magnetic moment of a transition metal at
an interface is mainly due to the suppression of the quenching of
the orbital component of the magnetization \cite{lic1}. In bulk
magnetic transition metals the orbital component of the
magnetization is strongly suppressed by the cubic crystal field.
In contrast at an interface the crystal symmetry is broken and the
quenching is only partial. This leads also to an enhancement of
the spin component which is strongly spin-orbit coupled to the
orbital one. Finally, for F/S interfaces a decrease of about 10\%
in the average magnetic moment in Fe has been reported in Fe/Nb
bilayers while cooling the sample through the superconducting
critical temperature \onlinecite{muhge3}. This evidence has been
explained \cite{bergeret} by the presence of a cryptoferromagnetic
state within islands of reduced exchange field in the Fe layer.
The phenomenon of cryptoferromagnetism \cite{anderson} consists of
the formation of a small-scale domain structure within a
ferromagnet in the vicinity of a F/S interface. In general,
however, the cryptoferromagnetic state in both samples is possible
only in the case of weak ferromagnets, such as Gd \cite{bergeret}.
No such behavior has yet been predicted or observed for Co.

\begin{figure}[h]
\narrowtext
\epsfysize=6cm
\epsfxsize=8cm
\centerline{\epsffile{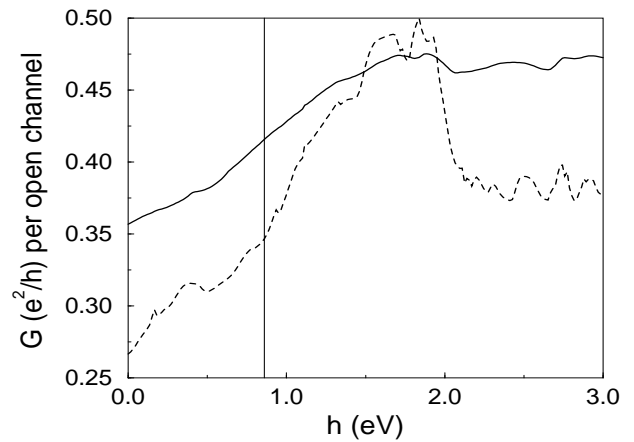}}
\caption{\it \label{G_vs_h0} $G_N$ (solid line) and $G_S$ (dashed line) as a function of the exchange field
in the Co monolayer at the interface, for fixed bulk exchange field.
The vertical line corresponds to the bulk exchange field in Co.}
\end{figure}

\begin{figure}[h]
\narrowtext
\epsfysize=6cm
\epsfxsize=8cm
\centerline{\epsffile{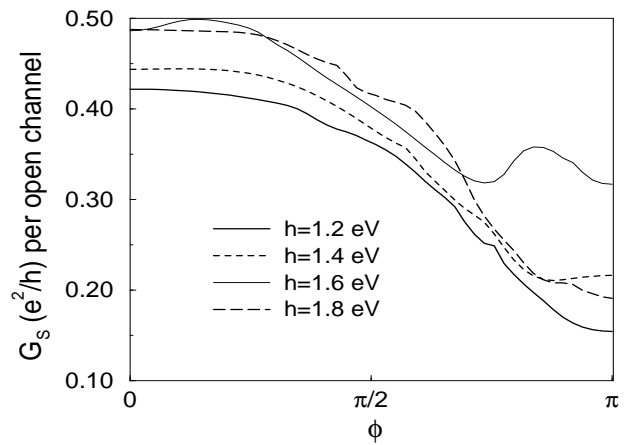}}
\caption{\it \label{G_vs_phi} $G_S$ as a function of the angle between the magnetization in the Co lead and
in the Co monolayer at the interface, for different values of the exchange
field $h$ in the monolayer.}
\end{figure}

\begin{figure}[h]
\narrowtext
\epsfysize=6cm
\epsfxsize=8cm
\centerline{\epsffile{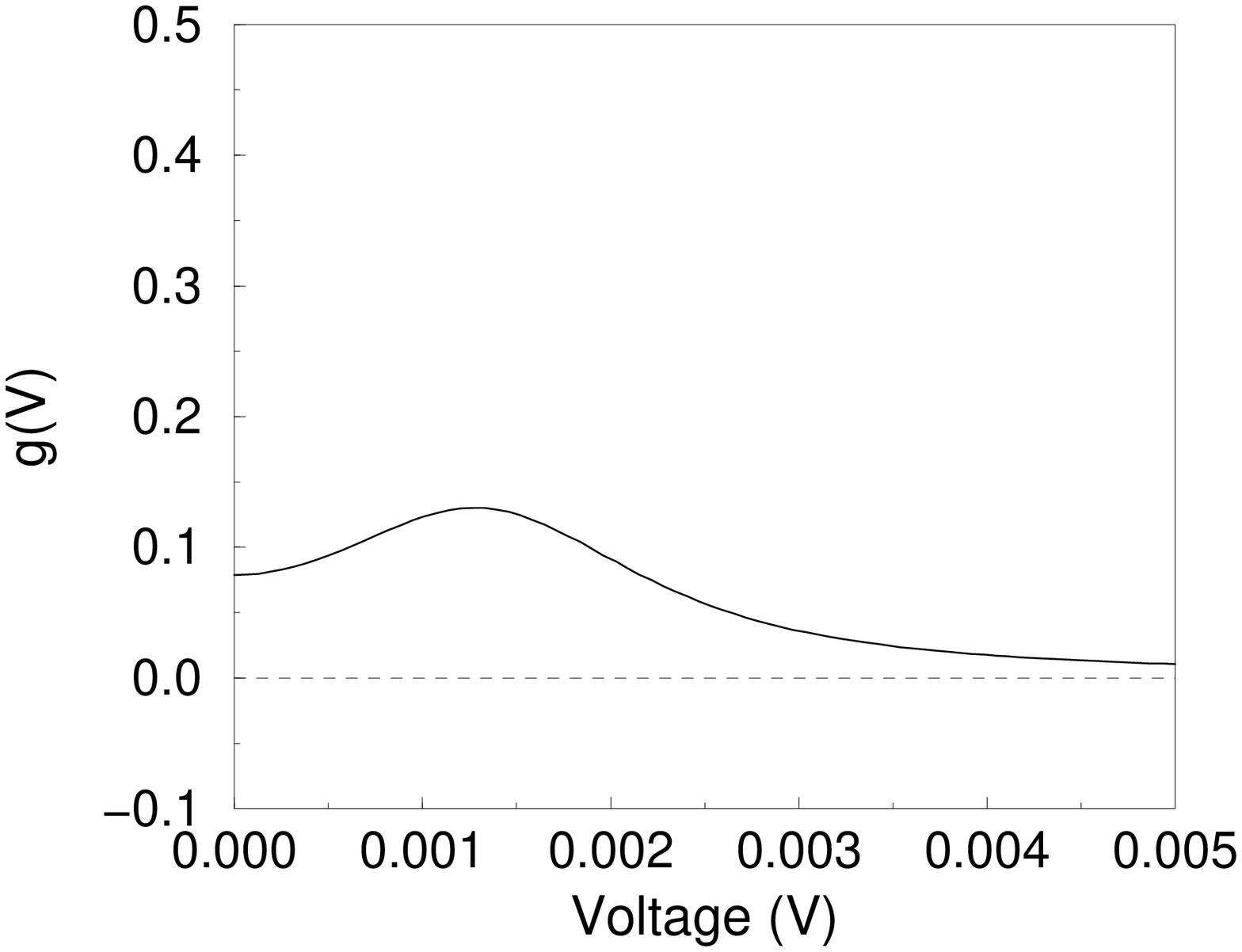}}
\caption{\it \label{g-Co/Pb} $g(V)$ curve for Co/Pb obtained from the spd-model when the exchange field in
the Co monolayer at the interface is $h=1.6$ eV.}
\end{figure}

In the case of references \onlinecite{Upd1,Upd2}, an altered
surface magnetization in Co could be produced by the peculiar
geometry of the sample: the portion of Co in contact with Pb has
an approximate area of (10 nm)$^2$, in which one domain can fit.
This suggests that the exchange field at the interface can be
larger than the exchange field in the bulk. In what follows, we
show that an increased exchange field at the interface does indeed
yield good agreement with the experimental $g(V)$ curve. Fig.
\ref{G_vs_h0} shows the calculated zero bias, zero temperature
conductances $G_S$ and $G_N$, for the Co/Pb junction in the
presence of a single surface monolayer of Co with arbitrary
exchange field $h$ entering all the orbitals. This shows that
there exists a range of values of the surface magnetization $h$
for which $G_S>G_N$ (in agreement with experiment), with the
largest $g(0)$ found for $h=1.84$ eV. We also considered the
possibility of a tilted magnetization in the inserted monolayer
with respect to the bulk magnetization, but as a general feature,
we find that for small angles $\phi$, $G_S$ does not vary much and
thereafter it decreases (see Fig. \ref{G_vs_phi}). In Fig.
\ref{g-Co/Pb} the $g(V)$ curve for Co/Pb is shown at 4.2 K for
$h=1.6$ eV, parallel to the bulk magnetization. This closely
matches the experimental plot in reference \onlinecite{Upd2} and
demonstrates that the experimental results for $g(V)$ of both
Cu/Pb and Co/Pb junctions can be reproduced, provided one accounts
for additional surface magnetism in the Co.


Finally we conclude by discussing the difference between the
polarization $\overline P$ of a bulk ferromagnet and the
polarization $P$ of a F/N interface made from the same
ferromagnet. Since the dependence of the transmission coefficients
on the energy is small (around 1 \%) in the range we are
considering here, we focus on the zero bias, zero temperature
limit. For a F/N interface the definition of $P$ given by the
equation (\ref{pol}) becomes
\begin{equation}
P= \frac{G^\uparrow -
G^\downarrow}{G^\uparrow + G^\downarrow} \; ,
\label{pol3}
\end{equation}
where $G^\uparrow(G^\downarrow)$ is the conductance for majority
(minority) electrons in units of $\frac{e^2}{h}$ normalized to the
corresponding number of open channels. From the results obtained
using the spd-model we find that while the polarization of bulk Co
is negative ($\overline{P}_{Co}=-0.400$), the polarization of a
Co/Pb(N) junction is positive (for instance, $P_{Co/Pb}=+0.400$
when $h$ at the interface equals the bulk value and
$P_{Co/Pb}=+0.275$ when at the interface $h = 1.6$ eV). This
striking difference occurs, because in the former case P is
determined solely by the DOS, with the minority electrons
possessing a larger DOS (mainly d-like) than the majority
electrons DOS (s,-p,-d-like) \cite{Ste}. In contrast for a
Co/Pb(N) junction, $P$ is also determined by the mismatch between
the band structures of the two materials. In this case, despite
their large DOS, minority electrons of Co are more strongly
scattered at the interface with Pb (whose DOS is mainly s- and
p-like) than Co majority electrons. This makes clear that in
general the polarization of a F/N junction also depends strongly
on the band structure of the non-magnetic material. As a further
example, we have also computed transport properties across an
Ir-Co interface. Assuming bulk magnetization at the interface we
obtain $P_{Co/Ir}=-0.010$, which has the opposite sign with
respect to Co/Pb. This arises since the DOS at the Fermi energy of
Ir is mainly d-like and the mismatch of the band structures with
Co is larger for majority than for minority electrons.

In conclusion we have shown that a detailed description of the
band structures of the individual materials and of the interface
is needed to accurately describe the I-V curves of S/F ballistic
junctions. In particular we have demonstrated that band structure
mismatch of the two materials can give rise to a polarization of
the whole junction which is completely different from the bulk
polarization of the  ferromagnetic. This casts some doubt on the
reliability of simple models based solely on surface scattering to
describe such junctions. Finally we found that for Co/Pb junctions
the experimental I-V curves are well reproduced if an enhancement
of the magnetization of Co at the interface is assumed. This is
consistent with the reduction of the quenching of the orbital
component of the magnetic moment of a ferromagnetic transition
metal at an interface, as reported recently in literature.

\end{multicols}

\end{document}